\documentclass[aip,jcp,preprint,amsmath,amsfonts,amssymb,showkeys]{revtex4-1}
\usepackage{graphicx,dcolumn,bm,xcolor,microtype}
\usepackage{tikz}
\usepackage{pgfplots}
\usepackage[version=4]{mhchem}

\newcommand{\alert}[1]{#1}
\newcommand{\cc}[1]{\multicolumn{1}{c}{#1}}
\newcommand{\mc}{\multicolumn}
\newcommand{\dfoot}{\cc{---\footnotemark[1]}}
\newcommand{\oldpack}{\textsc{Chem1D}}
\newcommand{\LegLag}{\textsc{LegLag}}

\newcolumntype{d}{D{.}{\cdot}{6}}
\makeatletter
\def\DC@endright{$\hfil\egroup\@dcolcolor\box\z@\box\tw@\dcolreset}
\def\dcolcolor#1{\gdef\@dcolcolor{\color{black}}}
\def\dcolreset{\dcolcolor{black}}
\dcolcolor{black}
\makeatother

\newcommand{\eps}{\epsilon}
\newcommand{\G}{\Gamma}
\newcommand{\om}{\omega}
\newcommand{\bD}{\mathbf{D}}
\newcommand{\bF}{\mathbf{F}}
\newcommand{\bL}{\mathbf{L}}
\newcommand{\bM}{\mathbf{M}}
\newcommand{\bR}{\mathbf{R}}

\newcommand{\cL}{\mathcal{L}}
\newcommand{\cR}{\mathcal{R}}
\newcommand{\cM}{\mathcal{M}}
\newcommand{\EHF}{$E_\mathrm{HF}$}
\newcommand{\EMP}{$E_\mathrm{MP2}$}
\newcommand{\EMPP}{$E_\mathrm{MP3}$}
\newcommand{\Eatom}{$E_{\rm atom}$}
\newcommand{\EAB}{$E_{\rm AB}$}
\newcommand{\EBC}{$E_{\rm BC}$}
\newcommand{\ECD}{$E_{\rm CD}$}
\newcommand{\RAB}{$R_{\rm AB}$}
\newcommand{\RBA}{$R_{\rm BA}$}
\newcommand{\RBC}{$R_{\rm BC}$}
\newcommand{\Eh}{$E_{\rm h}$}

\definecolor{mred}{rgb}{0.790588, 0.201176, 0.}
\definecolor{mgreen}{rgb}{0., 0.596078, 0.109804}
\definecolor{mblue}{rgb}{0.192157, 0.388235, 0.807843}
\definecolor{myellow}{rgb}{1., 0.607843, 0.}

\begin{document}

\title{Molecular electronic structure in \alert{one-dimensional Coulomb systems}}

\author{Caleb J. Ball}
\thanks{Corresponding author}
\email{caleb.ball@anu.edu.au}
\affiliation{Research School of Chemistry, Australian National University, Canberra ACT 2601, Australia}
\author{Pierre-Fran{\c c}ois Loos}
\email{pf.loos@anu.edu.au}
\affiliation{Research School of Chemistry, Australian National University, Canberra ACT 2601, Australia}
\affiliation{Laboratoire de Chimie et Physique Quantiques, Universit\'e de Toulouse, CNRS, UPS, France}
\author{Peter M. W. Gill}
\email{peter.gill@anu.edu.au}
\affiliation{Research School of Chemistry, Australian National University, Canberra ACT 2601, Australia}

\begin{abstract}
Following two recent papers [Phys. Chem. Chem. Phys. 2015, \textbf{17}, 3196; Mol. Phys. 2015, \textbf{113}, 1843], we perform a larger-scale study of chemical structure in one dimension (1D).
We identify a wide, and occasionally surprising, variety of stable 1D compounds (from diatomics to tetra-atomics) as well as a small collection of stable polymeric structures. 
We define the exclusion potential, a 1D analogue of the electrostatic potential, and show that it can be used to rationalise the nature of bonding within molecules.
This allows us to construct a small set of simple rules which can predict whether a putative 1D molecule should be stable.
\end{abstract}

\maketitle

\section{\label{sec:introduction} Introduction}
Recently, we introduced \oldpack, a program for electronic structure calculations on one-dimensional (1D) molecules. \cite{ball15, loos15} 
Unlike previous workers who used softened\cite{wagner12, stoudenmire12} or otherwise altered\cite{rosenthal71, Doren85, Doren87, Loeser85, Herschbach86, Loeser86a, Loeser86b}  interelectronic interactions in their studies of 1D chemical systems, \oldpack\ employs the unadorned Coulomb operator $|x|^{-1}$.
This potential introduces a non-integrable singularity which requires special treatment. 
Building on compelling arguments from the mathematical physics community, \cite{Oliveira09, Oliveira10, NunezYepez11, Oliveira12, NunezYepez14} our program avoids Coulombic divergences by solving the Schr\"odinger equation with Dirichlet boundary conditions that require the wavefunction to vanish wherever two particles---electrons or nuclei---touch.

The Dirichlet conditions have three chemically interesting consequences.
First, that molecular energies are spin-blind, i.e.~they are invariant with respect to spin-flips.
Second, that a ``super-Pauli'' exclusion rule applies, i.e.~an orbital cannot be occupied by more than one electron.
Third, that the nuclei become impenetrable, i.e.~electrons are unable to tunnel through them. \cite{Lee11a, Astrakharchik11, QR12, 1DEG13, Ringium13, Wirium14, HF1DEG16}

\alert{
The severity of these effects means that this model does not reflect the same type of experimental systems as the ``quasi-1D'' methods characterised by softened Coulomb interactions, which permit the nuclei to be penetrable and electrons to pair within spatial orbitals.
This reflects situations where the 1D confinement is not strict, and so they are well suited to modelling confined experimental systems such as ultracold atoms confined within a 1D trap.\cite{paredes04,moritz05,haller09}
}

\alert{
In contrast, the Coulomb interaction used in this work describes particles which are \emph{strictly} restricted to move within a 1D sub-space of three-dimensional space.
Early models of 1D atoms using this interaction have been used to study the effects of external fields upon Rydberg atoms\cite{Burnett93, Mayle07} and the dynamics of surface-state electrons in liquid helium. \cite{Nieto00, Patil01}
This description of 1D chemistry also} has interesting connections with the exotic chemistry of ultra-high magnetic fields (such as those in white dwarf stars), where the electronic cloud is dramatically compressed perpendicular to the magnetic field.\cite{Schmelcher90, Lange12, Schmelcher12}
In these extreme conditions, where magnetic effects compete with Coulombic forces, entirely new bonding paradigms emerge. \cite{Schmelcher90, Schmelcher97, Tellgren08, Tellgren09, Lange12, Schmelcher12, Boblest14, Stopkowicz15}

Unfortunately, our previous investigation \cite{ball15} of 1D chemistry suffered from debilitating numerical stability issues. 
The \oldpack\ program uses basis sets related to the exact wavefunctions \cite{loos15} of the hydrogen molecule cation \ce{H2+} but these quickly develop near-linear-dependence problems that prevent \oldpack\ from achieving basis set convergence even for relatively modest molecular systems.

In this paper we describe \LegLag, a more numerically stable version of \oldpack, which can be applied to a wider range of molecules to gain deeper insight into 1D chemistry.
In Sec.~\ref{sec:theory} we introduce the orthogonal set of basis functions which \LegLag\ employs and then briefly discuss the structure of the program. 
In Sec.~\ref{sec:results} we undertake an extensive study of 1D molecules, identifying multiple classes of stable species and the factors which lead to their stability. 
Finally, in Sec.~\ref{sec:rules}, we present a set of rules that govern chemical bonding in 1D. 
Unless otherwise stated, atomic units are used throughout.

\begin{figure}
	\includegraphics[width=0.45\textwidth]{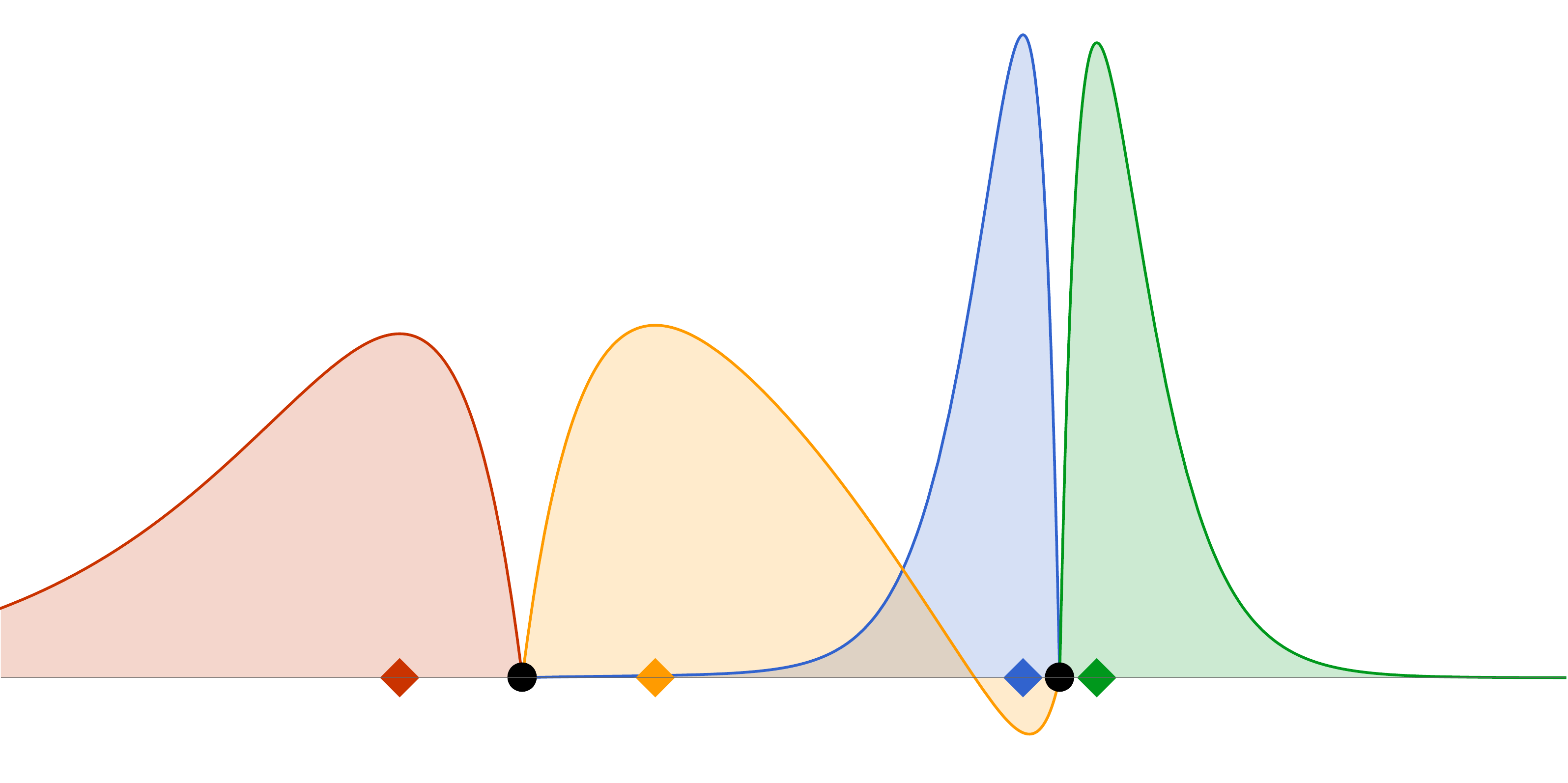}
	\caption{
	\label{fig:orbital_example}
	The ground state of the \ce{HLi} molecule. 
	Black circles represent the nuclei. 
	Each coloured region represents a singly-occupied orbital and the corresponding coloured diamond shows the most likely position of the electron.}
\end{figure}

\section{\label{sec:theory} Theory and Implementation}
Under Dirichlet boundary conditions, nuclei are impenetrable to electrons and each electron in a molecule is therefore confined to a ray or line segment by the nuclei closest to it.  In this way, the $M$ nuclei divide 1D space into two semi-infinite domains and $M-1$ finite domains.  Each domain supports a set of orbitals that vanish at the boundaries of the domain and outside it.
Figure \ref{fig:orbital_example} illustrates this for a small diatomic molecule.

In order to specify the constitution of a molecule, we employ a notation in which atomic symbols indicate nuclei and subscripts indicate the numbers of electrons in the intervening domains.
For example, \ce{_1Li4B3H1} is a triatomic with a lithium, boron and hydrogen nucleus arranged from left to right in that order. 
There is one electron to the left of the lithium nucleus, four electrons between the lithium and the boron nuclei, three electrons between the boron and the hydrogen nuclei and one electron to the right of the hydrogen nucleus.
In the present work, we consider only ground states and assume that the $n$ electrons within a domain singly occupy the $n$ lowest-energy orbitals.

\begin{figure*}
	\includegraphics[width=0.9\textwidth]{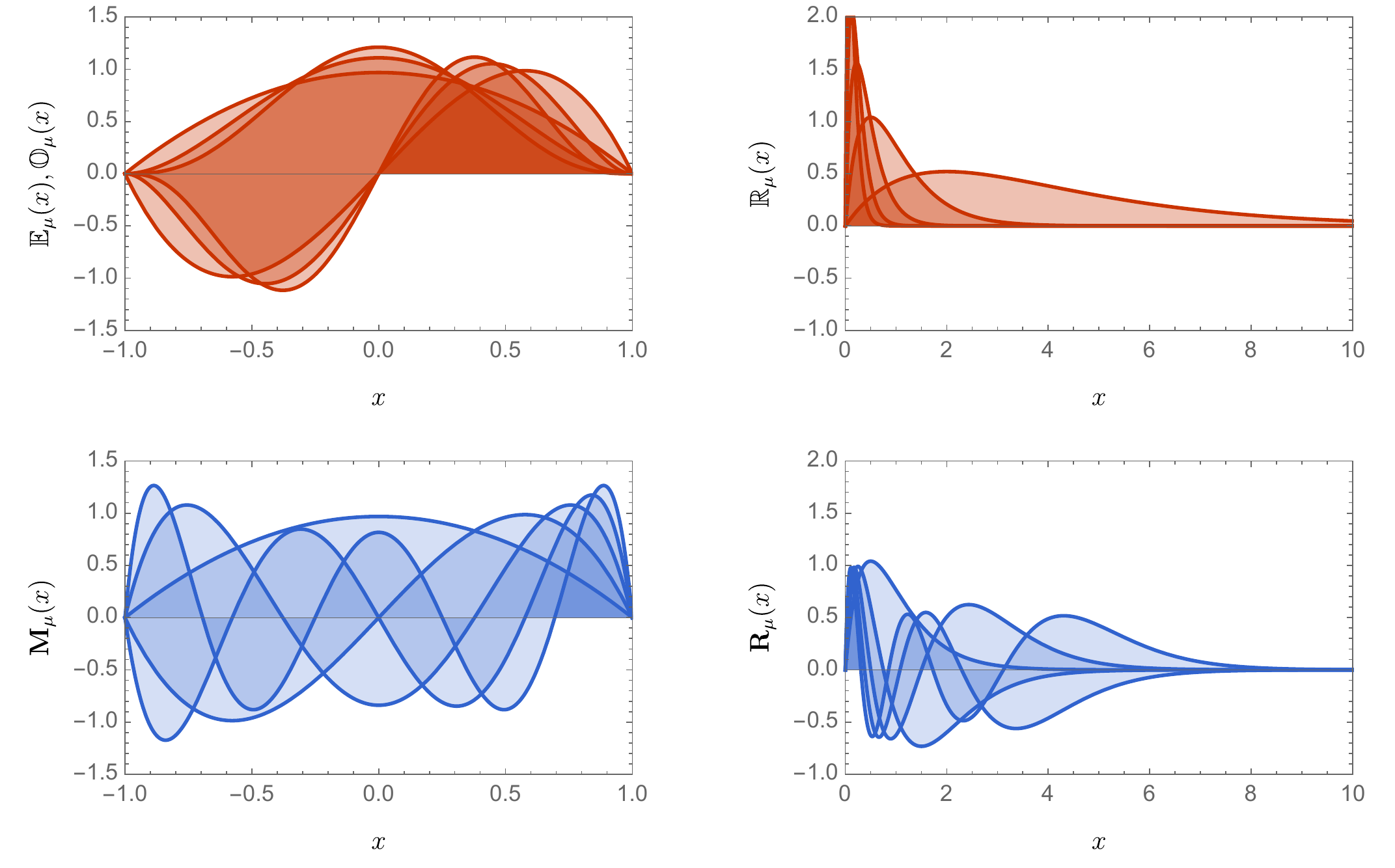}
	\caption{
	\label{fig:bf_comparison}
	A comparison of the basis functions used in \oldpack\ (top row, in red) with those used in \LegLag\ (bottom row, in blue). 
	A finite domain is shown on the left and an infinite domain on the right.}
\end{figure*}

\subsection{\label{subsec:basis} Basis sets}
There are three types of domain -- left, right and middle -- and we require a set of basis functions for each.
The functions should vanish at the domain boundaries and form a complete set.

\oldpack\ uses the functions
\begin{subequations} \label{eq:chem1dbasis}
\begin{align}
	\mathbb{L}_\mu(s)	& = 2\mu^3\alpha^{1/2} s \exp(-\mu^2 s)							\label{eq:lbf}	\\
	\mathbb{R}_\mu(t)	& = 2\mu^3\alpha^{1/2} t \exp(-\mu^2 t) 						\label{eq:rbf}	\\
	\mathbb{E}_\mu(z)	& = \sqrt{\frac{(2\mu+1)_{1/2}}{\om\pi^{1/2}}}	\ (1-z^2)^\mu	\label{eq:ebf}	\\
	\mathbb{O}_\mu(z)	& = \sqrt{\frac{2(2\mu+1)_{3/2}}{\om\pi^{1/2}}}	\ z(1-z^2)^\mu	\label{eq:obf}
\end{align}
\end{subequations}
where $s = \alpha(A-x)$, $t = \alpha(x-B)$ and $z = (x-C)/\om$ are the reduced coordinates in the left, right and middle domains respectively, $A$ and $B$ are the positions of the leftmost and rightmost nuclei, and $C$ and $\om$ are the center and halfwidth of a middle domain, $\alpha > 0$ is an exponent, and $(a)_n$ is the Pochhammer symbol. \cite{NISTbook} 
The $\mathbb{L}_\mu$ and $\mathbb{R}_\mu$ functions are used in the left and right domains, respectively.  $\mathbb{E}_\mu$ and $\mathbb{O}_\mu$ are used in the middle domains.

One disadvantage of these functions is their increasing linear dependence as the size of the basis set grows, which creates numerical instability in the orthogonalisation step of the Pople-Nesbet Hartree-Fock (HF) method. \cite{SzaboBook}
This limits the size of basis set that can be employed before unacceptable numerical precision is lost.

A second disadvantage of this basis set is that, because the $\mathbb{E}_\mu$ and $\mathbb{O}_\mu$ functions are increasingly peaked around the middle of the domain, they struggle to describe details of the molecular orbitals near domain boundaries.  This becomes particularly problematic when the domain contains more than one electron. \cite{ball15}

In contrast, \LegLag\ uses the basis functions
\begin{subequations} \label{eq:LRM-bf}
\begin{align}
	\bL_\mu(s)	& = \sqrt{\frac{8 \alpha}{(\mu)_2}}		\ s L_{\mu-1}^2 (2s) \exp(-s)	\label{eq:L-bf}	\\
	\bR_\mu(t)	& = \sqrt{\frac{8 \alpha}{(\mu)_2}}		\ t L_{\mu-1}^2 (2t) \exp(-t)	\label{eq:R-bf}	\\
	\bM_\mu(z)	& = \sqrt{\frac{\mu+3/2}{\om(\mu)_4}}	\ P_{\mu+1}^2(z)				\label{eq:M-bf}
\end{align}
\end{subequations}
where $L_m^2$ and $P_m^2$ are second-order associated Laguerre and Legendre polynomials. \cite{NISTbook}
(Our package's name stems from its use of Legendre and Laguerre polynomials.)
The $\bL_\mu$, $\bR_\mu$ and $\bM_\mu$ functions are used in the left, right and middle domains, respectively, and are mutually orthogonal.  The $\bM_\mu$ are evenly distributed across the domain, as Fig.~\ref{fig:bf_comparison} shows.

\subsection{\label{subsec:integrals} Integrals}
In earlier work, we discovered \cite{ball15} that HF calculations \cite{SzaboBook} give unexpectedly accurate results in 1D. 
It also appears, in contrast to the situation in 3D,\cite{Deceptive86, SlowUMP88, O291} that the M{\o}ller-Plesset (MP) perturbation series \cite{SzaboBook} in 1D often converges rapidly to the exact energy. \cite{loos15}  However, to perform such calculations it is necessary to evaluate the integrals
\begin{subequations} \label{eq:ints}
\begin{align}
	(\bF_\mu | \bF_\nu)							& = \int \bD_{\mu\nu}(x) \,dx = \delta_{\mu\nu}								\\
	(\bF_\mu | \hat{T} | \bF_\nu)				& = \frac{1}{2} \int \bF_\mu^\prime(x) \bF_\nu^\prime(x) \,dx				\\	\label{eq:potentialint}
	(\bF_\mu | \hat{V} | \bF_\nu)				& = \int \frac{\bD_{\mu\nu}(y) }{|x - y|} \,dy								\\	\label{eq:trueeri}
	(\bF_\mu \bF_\nu | \bF_\lambda \bF_\sigma)	& = \iint \frac{\bD_{\mu\nu}(x) \bD_{\lambda\sigma}(y)  }{|x - y|} \,dx\,dy
\end{align}
\end{subequations}
where $\bF \in \{\bL, \bM, \bR\}$, $\bD_{\mu\nu}(x) = \bF_\mu(x) \bF_\nu(x)$ is a density component, $\hat{T} = -\nabla^2/2$ is the kinetic energy operator and $\delta_{\mu\nu}$ is the Kronecker delta function. \cite{NISTbook}

If the four basis functions are in the same domain, the singularity of the Coulomb operator causes $(\bF_\mu \bF_\nu | \bF_\lambda \bF_\sigma)$ to diverge.  However, the antisymmetrized integral
\begin{equation}
	(\bF_\mu \bF_\nu | |\bF_\lambda \bF_\sigma) = (\bF_\mu \bF_\nu | \bF_\lambda \bF_\sigma) - (\bF_\mu \bF_\sigma | \bF_\lambda \bF_\nu)
\end{equation}
is finite and can be found from quasi-integrals \cite{ball15} using
\begin{equation}
	(\bF_\mu \bF_\nu || \bF_\lambda \bF_\sigma) = \{\bF_\mu \bF_\nu | \bF_\lambda \bF_\sigma\} - \{\bF_\mu \bF_\sigma | \bF_\lambda \bF_\nu\}
\end{equation}

Because $\bR_\mu$ is the image of $\bL_\mu$ under inversion through the molecular mid-point, formulae involving only $\bR_\mu$ and $\bM_\mu$ can be found from the equivalent formulae involving $\bL_\mu$ and $\bM_\mu$.  We will therefore not discuss the former.

Integral formulae are given in Appendix \ref{app:ints} and most of the necessary special functions are evaluated by calling external libraries.  However, because we invariably need a range of values for the $a$ and $b$ parameters in the Tricomi confluent hypergeometric functions \cite{NISTbook} $U(a, b, z)$ required for Coulomb integrals involving the $\bL_\mu$ functions, it is more efficient to compute these functions recursively.  It has been shown that backwards recurrence in the $a$ parameter is numerically stable and our algorithm exploits this.\cite{deano08}  To obtain the starting values for this recurrence we use an asymptotic expansion that is valid when $2a - b$ is large and positive.\cite{abad97, abad99}  Our numerical experiments have shown that for arguments $z > 10$ this expansion converges at an unacceptable rate.  Our algorithm therefore uses Miller's method\cite{bickley52, NISTbook} when $z > 8$.

We detect and avoid computing negligible integrals using the Coulomb upper bound\cite{Bound94}
\begin{equation} \label{eq:bound}
	\left| (P|Q) \right| \le \min(V_P^* S_Q^* , S_P^* V_Q^*)
\end{equation}
where $V_P^*$ is the maximum potential of $P(x)$ in the domain of $Q(x)$ and $S_Q^*$ is the integral of $|Q(x)|$.  The $V^*$ and $S^*$ values can be found using expressions in Section \ref{app:prop}.

\subsection{\label{subsec:implementation} Implementation}
Aside from integral evaluation, \LegLag~closely follows the algorithms employed in \oldpack. 
For a comprehensive description of these, see the paper by Ball and Gill. \cite{ball15}

\LegLag~has been implemented using the Python programming language (version 3.41) in combination with the Cython language extension for compute-intensive bottlenecks. 
It employs the external Numpy library for data structures and linear algebra operations and the Scipy library for computing some of the special functions.

A significant feature of \LegLag\ is that it can be easily controlled by external Python scripts. 
In generating the data presented in Sec.~\ref{sec:results}, we have made extensive use of scripts that use the numerical function minimiser available in Scipy to optimize molecular geometries.

\begin{figure}
	\includegraphics[width=0.45\textwidth]{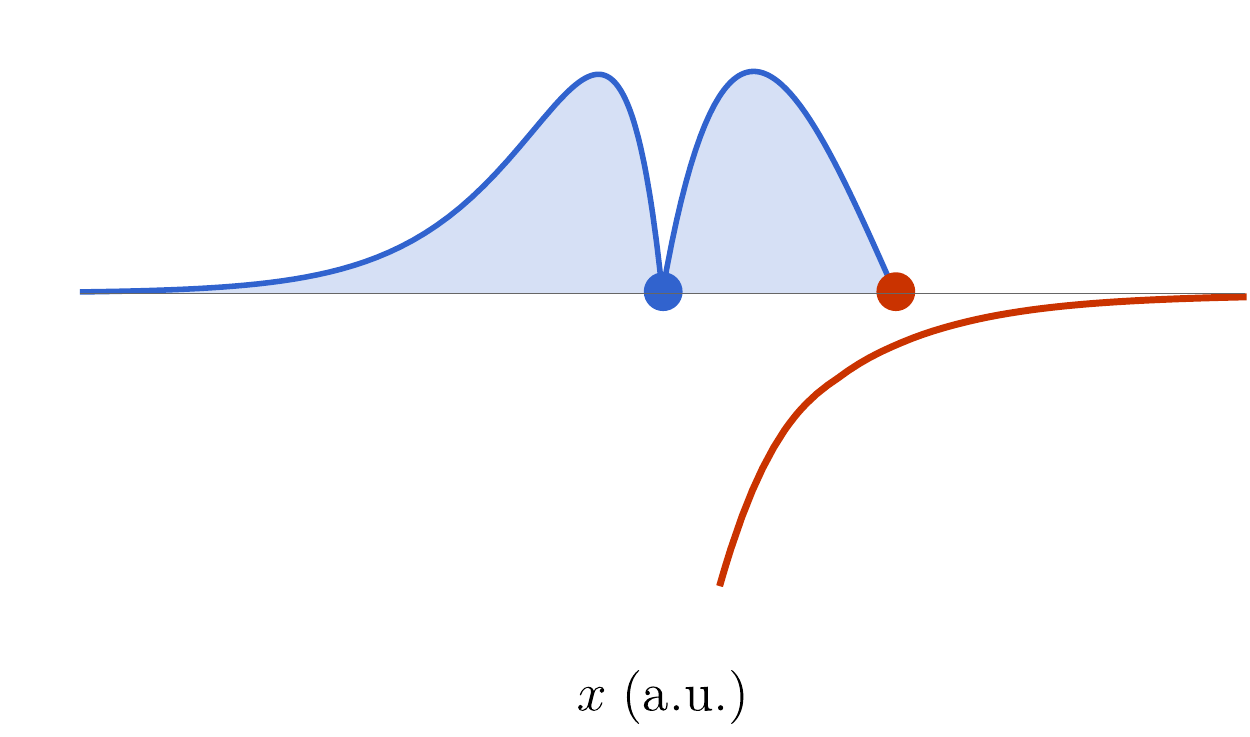}
	\caption{
	\label{fig:potential_construction}
	The exclusion potential (red) of a \ce{_1He_1} atom (blue). The blue regions show the occupied orbitals when computing the exclusion potential at the position of the red dot.}
\end{figure}

\subsection{\label{subsec:potential} Exclusion potential}
In 3D, the molecular electrostatic potential\cite{ESP93} (MESP) is the limit of the ratio of the Coulomb energy of a test particle to the magnitude of its charge, as that charge approaches zero.
It is a potent tool for understanding chemical behaviour and can reveal, for example, electrophilic or nucleophilic regions.
Unfortunately, however, the MESP diverges at all points in a 1D system except where the electron density vanishes. \cite{Ringium13}
Therefore, to define a meaningful potential in a 1D molecule, we must insist that the test particle create a new Dirichlet node at its position.
We call the resulting potential the ``exclusion potential'' to emphasise that, in contradistinction to the 3D analog, the test particle in 1D excludes electrons from its neighbourhood and thereby significantly perturbs the system.

Figure~\ref{fig:potential_construction} shows the exclusion potential for a \ce{_1He_1} atom as well as the perturbed orbitals for a given position of the test particle.
Note how the Dirichlet node created by the test particle compresses the right orbital, and prevents the electron which occupies it from extending to the right.

\begin{table*}
\caption{Total energies (\Eh), ionization energies and electron affinities (eV) of 1D atoms using the (30,50) basis set.}
\label{tab:atoms}
\begin{ruledtabular}
\begin{tabular}{@{}l *{3}{D{.}{.}{3.6}} *{3}{D{.}{.}{2.3}} *{3}{D{.}{.}{1.3}}}
	Atom	&			\mc{3}{c}{Total energies}				&		\mc{3}{c}{Ionization energies}			&	\mc{3}{c}{Electron affinities\footnotemark[1]}	\\
								\cline{2-4}											\cline{5-7}										\cline{8-10}
			&	\cc{--\EHF}	&	\cc{--\EMP}	&	\cc{--\EMPP}	&	\cc{HF}		&	\cc{MP2}	&	\cc{MP3}	&	\cc{HF}		&	\cc{MP2}		&	\cc{MP3}	\\
	\colrule
	H		&	0.500000	&	0.500000	&	0.500000		&	13.606		&	13.606		&	13.606		&	3.893		&	3.939			&	3.961		\\
	He		&	3.242922	&	3.244986	&	3.245611		&	33.822		&	33.878		&	33.895		&	\cc{---}		&	\cc{---}			&	\cc{---}		\\
	Li		&	8.007756	&	8.01112		&	8.01179		&	4.486		&	4.517		&	4.522		&	1.395		&	1.410			&	1.414		\\
	Be		&	15.415912	&	15.4226	&	15.4236		&	10.348		&	10.400		&	10.408		&	\cc{---}		&	\cc{---}			&	\cc{---}		\\
	B		&	25.35751	&	25.3671	&	25.3684		&	2.068		&	2.09		&	2.099		&	0.64		&	0.65			&	0.65		\\
	C		&	38.09038	&	38.105		&	38.107			&	4.670		&	4.719		&	4.73		&	\cc{---}		&	\cc{---}			&	\cc{---}		\\
	N		&	53.569		&	53.59		&	53.6			&	1.1			&	1.1			&	1.1			&	0.3			&	0.3				&	0.3			\\
	O		&	71.9293	&	71.95		&	71.96			&	2.516		&	2.548		&	2.556		&	\cc{---}		&	\cc{---}			&	\cc{---}		\\
	F		&	93.1		&	93.2		&	93.2			&	0.5			&	0.5			&	0.5			&				&					&				\\
	Ne		&	117.31		&	117.35		&	117.35			&	1.5			&	1.5			&	1.5			&	\cc{---}		&	\cc{---}			&	\cc{---}		\\
\end{tabular}
\end{ruledtabular}
\footnotetext[1]{The electron affinities of \ce{He}, \ce{Be}, \ce{C}, \ce{O} and \ce{Ne} are omitted because the anions of these species are auto-ionising. \cite{ball15}}
\end{table*}

\section{\label{sec:results} Results and Discussion}
To begin to understand the nature of chemical bonding in 1D, we have performed an extensive search for stable molecules. 
After presenting accurate atomic energies, we will discuss the structures of a wide variety of small molecules and a small set of polymeric systems.

The Periodic Table in 1D has only two groups\cite{loos15} and we will frequently refer to alkalis (\ce{H}, \ce{Li}, \ce{B}, \ldots which have an odd number of electrons and a permanent dipole moment) and nobles (\ce{He}, \ce{Be}, \ce{C}, \ldots which have an even number of electrons, are symmetrical and have no dipole.)

All of the calculations that we report use 30 basis functions in each of the left and right domains and 50 functions in each of the finite domains.  We will refer to this as the (30,50) basis set.
We report only the digits that have converged as the basis set is increased to the (30,50) set.

\subsection{\label{subsec:atoms} Atoms}
Our first task was to choose the exponent $\alpha$ in $\bL_\mu(x)$ and $\bR_\mu(x)$ that yields the best energies as the basis set is increased. 
We expected that $\alpha$ would be determined largely by the innermost (and lowest-energy) orbitals, and that therefore $\alpha \approx Z$, where $Z$ is the nuclear charge of the atom in question.  We were therefore surprised to find that this is not the case and that, except for hydrogen, the optimal exponent is always close to $\alpha = 2$, a compromise that attempts to describe both the compact inner orbitals and the diffuse outer orbitals.  After this discovery, we used $\alpha = 2$ for all atoms.

In our first foray into 1D chemistry, \cite{loos15} we used multiple-precision arithmetic in \textsc{Mathematica}\cite{mathematica} to compute the near-exact HF, MP2 and MP3 energies, ionisation energies and electron affinities of the ground-state atoms up to \ce{$_5$Ne_5}.  Our subsequent (double-precision) \oldpack\ program was often unable to reproduce these energies, principally because of its inadequate basis functions \eqref{eq:chem1dbasis}.  Table \ref{tab:atoms} shows that our (double-precision) \LegLag\ calculations are much more successful in capturing the energies but the (30,50) basis still struggles for the largest atoms and, in particular, fails to yield any significant figures for the electron affinity of \ce{$_4$F_5}.

\subsection{\label{subsec:diatomics} Diatomics}
\begingroup
\squeezetable
\begin{table*}
\caption{
\label{tab:diatomics}
Equilibrium bond lengths (bohr), total energies (\Eh) and dissociation energies (m\Eh) of diatomic molecules.
}
\begin{ruledtabular}
\begin{tabular}{@{}l *{3}{D{.}{.}{2.3}} *{3}{D{.}{.}{3.6}} *{3}{D{.}{.}{3.6}}}
	 \cc{Molecule}	&		\mc{3}{c}{Bond length}		&			\mc{3}{c}{Total energy}				&		\mc{3}{c}{Dissociation energy}			\\
								\cline{2-4}									\cline{5-7}										\cline{8-10}
	\cc{AB}		& \cc{HF}	& \cc{MP2}	& \cc{MP3}	& \cc{--\EHF}	& \cc{--\EMP}	& \cc{--\EMPP}	&	\cc{HF}		&	\cc{MP2}	&	\cc{MP3}		\\
	\colrule
	\ \,\ce{H1H1}	&	2.636	&	2.637	&	2.638	&	1.184572	&	1.185418	&	1.185728	&	184.572		&	185.418		&	185.728		\\
	\ce{_1H1He1}	&	2.025	&	2.027	&	2.027	&	3.880313	&	3.882619	&	3.883301	&	137.39		&	137.633		&	137.691		\\
	\ \,\ce{H2Li2}	&	5.345	&	5.323	&	5.320	&	8.544163	&	8.547920	&	8.548659	&	36.407		&	36.800		&	36.871		\\
	\ce{_1H2Li1}	&	5.152	&	5.141	&	5.142	&	8.681782	&	8.686367	&	8.687589	&	174.025		&	175.25		&	175.80		\\
	\ce{_1H2Be2}	&	3.966	&	3.961	&	3.962	&	16.079548	&	16.08707	&	16.08845	&	163.636		&	164.50		&	164.85		\\
	\ce{_1H3B2}	    &	8.880	&	8.810	&	8.806	&	26.020047	&	26.0310		&	26.0329		&	 \dcolcolor{red}95.492		&	 \dcolcolor{red}96.1		&	 \dcolcolor{red}96.1		\\
	\ce{_1H2B3}		&	3.298	&	3.296	&	3.296	&	25.957601	&	25.96793	&	25.96949	&	100.093		&	100.85		&	101.13		\\
	\ \,\ce{H3B3}	&	10.349	&	10.238	&	10.235	&	25.863890	&	25.8736		&	25.8748		&	6.382		&	6.5			&	6.5			\\
	\ce{_1H3C3}		&	6.666	&	6.635	&	6.633	&	38.756672	&	38.7721		&	38.7745		&	166.290		&	167.4		&	167.9		\\
	\ce{_1H4N3}		&	14.316	&	14.268	&	14.257	&	54.22372	&	54.244		&	54.247		&	 \dcolcolor{red}52.470		&	 \dcolcolor{red}53			&	 \dcolcolor{red}53			\\
	\ce{_1H3N4}		&	5.407	&	5.392	&	5.392	&	54.218224	&	54.2379		&	54.2407		&	149.222		&	150.2		&	150.5		\\
	\ \,\ce{H4N4}	&	19.20	&	18.168	&	18.131	&	54.0703		&	54.089		&	54.091		&	1.3			&	1			&	1			\\
	\ce{_1H4O4}		&	10.468	&	10.383	&	10.378	&	72.590721	&	72.616		&	72.620		&	\dcolcolor{red}110.787		&	\dcolcolor{red}112			&	\dcolcolor{red}112			\\
	\ce{_1He2Li2}	&	4.606	&	4.586	&	4.584	&	11.260655	&	11.266223	&	11.267543	&	9.977		&	10.118		&	10.144		\\
	\ce{_1He3B3} 	&	11.174	&	11.170	&	11.003	&	28.600892	&	28.6126		&	28.6145		&	0.461		&	0.5			&	0.5			\\
	\ce{_1Li3Li2}	&	8.693	&	8.644	&	8.637	&	16.064647	&	16.07183	&	16.07326	&	49.134		&	49.59		&	49.69		\\
	\ce{_2Li3Be2}	&	7.050	&	7.000	&	6.996	&	23.452479	&	23.46286	&	23.46460	&	28.811		&	29.16		&	29.22		\\
	\ce{_2Li4B2}	&	13.330	&	13.228	&	13.157	&	33.418876	&	33.4323		&	33.4343		&	53.611		&	54.1		&	54.2		\\
	\ce{_1Li4B3}	&	14.007	&	13.999	&	13.778	&	33.379031	&	33.3922		&	33.3941		&	13.766		&	14.0		&	14.0		\\
	\ce{_2Li4C3}	&	10.435	&	10.358	&	10.336	&	46.140625	&	46.1588		&	46.1614		&	42.486		&	43.0		&	43			\\
	\ce{_2Li4N4}	&	8.956	&	8.892	&	8.884	&	61.588547	&	61.6112		&	61.6143		&	11.788		&	12.3		&	12.4		\\
	\ce{_2Li5N3}	&	19.552	&	19.258	&	19.229	&	61.63192	&	61.654		&	61.658		&	\dcolcolor{red}44.7		&	\dcolcolor{red}45			&	\dcolcolor{red}45			\\
	\ce{_1Li5N4}	&	21.546	&	21.092	&	21.099	&	61.5802		&	61.602		&	61.605		&	3.5			&	3			&	3			\\
	\ce{_2Li5O4}	&	14.943	&	14.769	&	17.748	&	79.987067	&	80.015		&	80.019		&	49.98		&	51			&	51			\\
	\ce{_2Be4B3}	&	12.566	&	12.566	&	12.381	&	40.776885	&	40.7932		&	40.7955		&	3.464		&	3.6			&	3.6			\\
	\ce{_2Be5N4}	&	20.571	&	19.884	&	19.869	&	68.9851		&	69.010		&	69.014		&	0.2			&	0			&	0			\\
	\ce{_2B5B3}		&	19.349	&	19.233	&	19.003	&	50.733908	&	50.753		&	50.756		&	18.891		&	19			&	19			\\
	\ce{_3B5C3}		&	16.040	&	16.009	&	15.779	&	63.457101	&	63.481		&	63.484		&	9.21		&	9			&	9			\\
	\ce{_3B6N3}		&	26.480	&	25.946	&	25.912	&	78.949		&	78.977		&	78.981		&	22			&	22			&	22			\\
	\ce{_2B6N4}		&	27.514	&	26.939	&	26.869	&	78.933		&	78.961		&	78.96		&	6			&	6			&	6			\\
	\ce{_3B6O4}		&	21.138	&	20.799	&	20.735	&	97.3018		&	97.34		&	97.34		&	14.9		&	15			&	15			\\
	\ce{_3C6N4}		&	22.880	&	24.906	&	24.801	&	91.660		&	91.69		&	91.70		&	1			&	1			&	1			\\
	\ce{_3N7N4}		&	30.301	&	31.892	&	33.989	&	107.14		&	107.18		&	107.19		&	10			&	10			&	10			\\
	\ce{_4N7O4}		&	29.583	&	28.780	&	28.727	&	125.50		&	125.54		&	125.55		&	0			&	0			&	0			\\
\end{tabular}
\end{ruledtabular}
\end{table*}
\endgroup

\begin{figure}
	\includegraphics[width=0.45\textwidth]{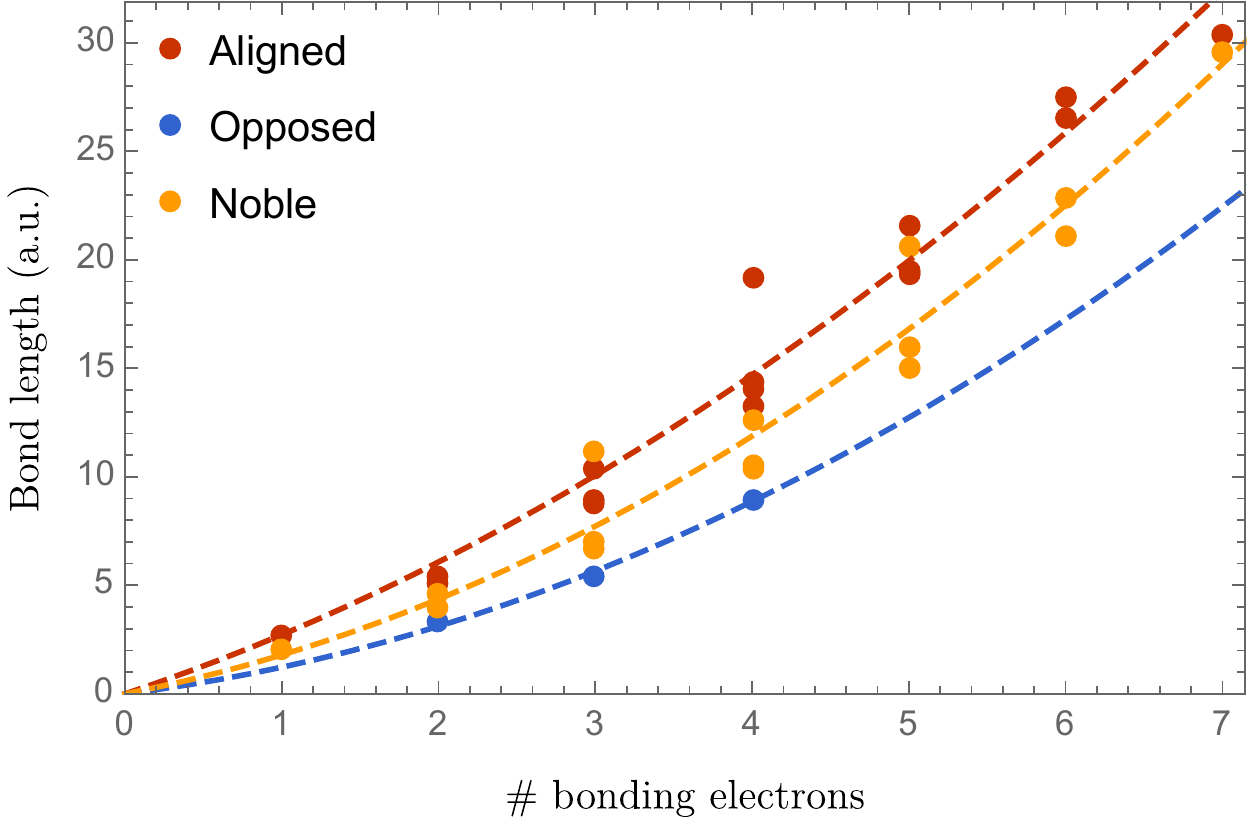}
	\caption{
	\label{fig:bondlengths}
	Variation of diatomic bond lengths (in bohrs) with the number of electrons in the middle domain. 
	Data are grouped according to the character of the molecule: aligned alkali-alkali, opposed alkali-alkali, and alkali-noble. 
	Quadratic least-square fits are shown as dotted lines.}
\end{figure}

Notwithstanding the deficiencies of the (30,50) basis for the largest atoms, \LegLag\ is able to treat a far wider variety of molecular systems than is possible in \oldpack\ and we have surveyed the diatomics with atoms up to oxygen and with all possible electronic configurations that can be generated from the ground-state atoms.  Table \ref{tab:diatomics} reports the bond lengths and energies of the diatomics that we have found to be stable, i.e.~lower in energy than their constituent atoms.  This set of results, which greatly extends our previous efforts, \cite{loos15, ball15} allows us significantly more insight into the mechanics of 1D bonding.

There appear to be four major factors that govern the binding between two atoms:
\begin{enumerate}
	\setlength\itemsep{0.5mm}
	\item Valence attraction to an alkali nucleus
	\item Nuclear shielding
	\item Dipole interactions
	\item Number of occupied domains
\end{enumerate}
and we now discuss each of these in turn, and pictorial representations can be seen in Figure~\ref{fig:bonding_diagrams}.

\begin{figure}
	\resizebox{0.45\textwidth}{!}{
	\include{fig5}
	}
	\caption{
	\label{fig:bonding_diagrams}
	1D analogues of Lewis dot diagrams representing the major factors governing diatomic bonding: a. valence shell interactions, b. nuclear shielding, c. dipole interactions and d. number of occupied domains.
	More stable configurations are represented in green, less stable in red.
	Dotted circles represent unoccupied orbitals.
	}
\end{figure}

Valence-nucleus attraction (Fig.~\ref{fig:bonding_diagrams}a) is strongest on the electron-deficient side of the alkali;  on the other side of an alkali, or on either side of a noble, such an interaction is shielded too effectively.  The four configurations of the HB molecule, viz.~\ce{_1H3B2}, \ce{_1H2B3}, \ce{H3B3} and \ce{H4B2}, illustrate this.  The first three are bound and, in each, at least one of the atoms presents its electron-deficient side to the other atom.  The fourth configuration, in which each atom presents its electron-rich side to the other, is unstable.

It follows that two nobles will not bind, as neither has an electron-deficient side.  In 3D, noble gas atoms can bind weakly through dispersion interactions, \cite{London30, London37, Pitzer55, StoneBook} but we have not seen evidence of this in 1D.  This may be an artefact of the (30,50) basis but we believe that such binding, if it exists, is likely to be very weak.

Nuclear shielding (Fig.~\ref{fig:bonding_diagrams}b) is also critical.  Lighter atoms bind more strongly because their nuclei are less shielded and this is true \emph{a fortiori} of the completely unshielded H atom.  As the shielding increases, binding energies drop rapidly, and bond strengths in nitrogen-containing molecules like \ce{_1Li5N4} and \ce{_3C6N4} are in the millihartree (m\Eh) range.

Dipole interactions (Fig.~\ref{fig:bonding_diagrams}c) also influence bond strengths.  The \ce{H1H1} and \ce{_1H1He1} molecules each hold a single electron in the internuclear domain and, naively, one might expect that \ce{_1H1He1} would be more strongly bound because of its shorter bond and greater attraction between the bonding electron and nuclei.  However, the bond strength in \ce{H1H1}, where the atomic dipoles are favourably aligned, exceeds that in \ce{_1H1He1} by roughly 50 m\Eh.  Similar arguments explain the relative strengths in \ce{H2Li2} and \ce{_1He2Li2}.  The rare instances, e.g.~\ce{_2Li4N4}, where a diatomic forms with opposed dipoles are driven by the attraction between two electron-deficient sides.

The number of occupied domains (Fig.~\ref{fig:bonding_diagrams}d) is also relevant, as the very different bond energies in the HB diatomics show. 
Why, for example, is \ce{H3B3} is so much more weakly bound than \ce{_1H3B2}, despite both having favourable dipole alignments?  The answer is that the six electrons are squeezed into two domains in \ce{H3B3}, rather than three in \ce{_1H3B2}.  To form the former from the latter, the electron from the left domain is promoted into a high-energy orbital in the right domain and this incurs a large energy cost.

\begin{figure*}
	\includegraphics[width=\linewidth]{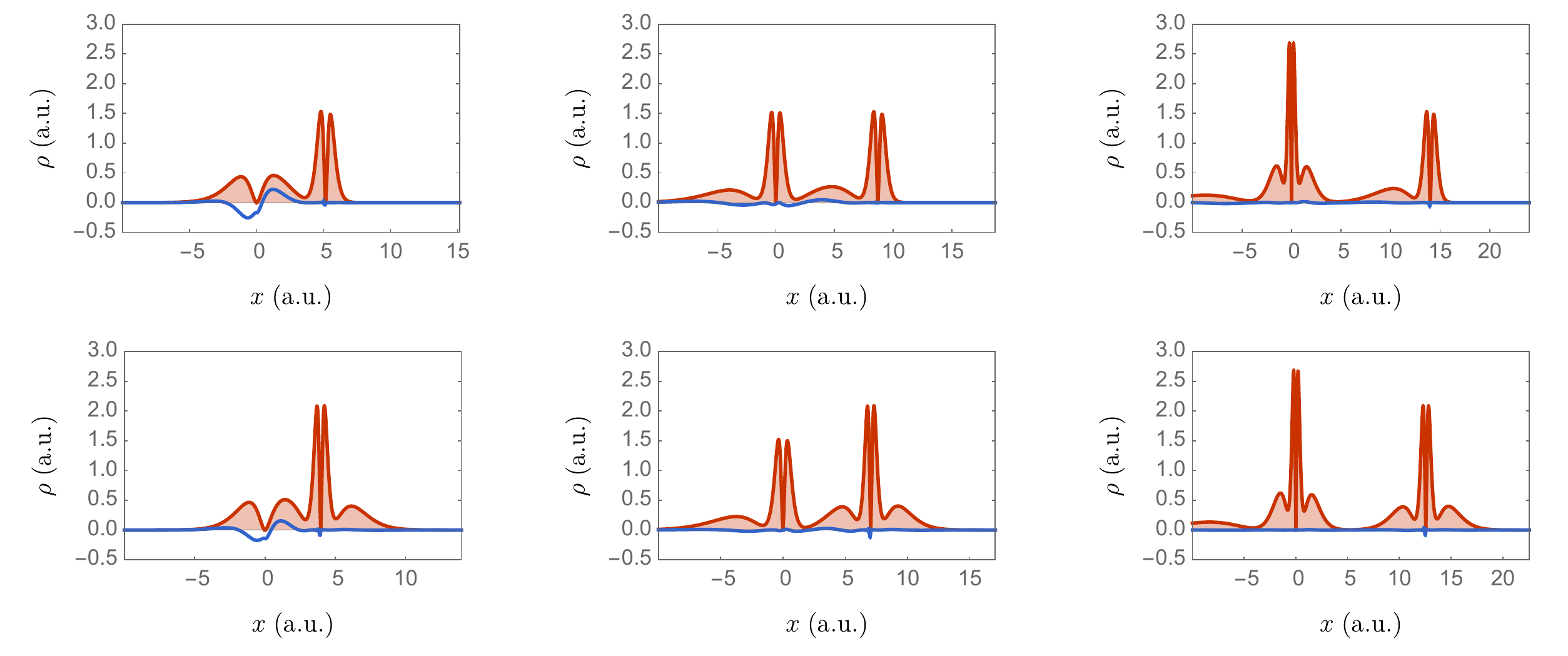}
	\caption{
	\label{fig:diatomic_densities}
	Electronic densities (red regions) and difference between molecular and atomic densities (blue lines) of six diatomic molecules. 
	In the top row these are \ce{_1H2Li1}, \ce{_2Li3Li1} and \ce{_3B4Li1}. 
	In the bottom these are \ce{_1H2Be2}, \ce{_2Li3Be2} and \ce{_3B4Be2}.}
\end{figure*}

In our earlier work,\cite{loos15, ball15} we discovered a few surprisingly long bonds.  However, the results in Table \ref{tab:diatomics} show that gargantuan bond lengths are not at all uncommon in 1D.
Figure~\ref{fig:bondlengths} reveals a strong correlation between the length of the middle domain of a diatomic and the number of electrons occupying it. 
If the data are grouped into those with aligned dipoles, those with opposed dipoles and those with a noble atom, strong parabolic trends emerge. 
Similar behaviour is found for estimates of atomic radii. \cite{loos15}

This similarity might suggest that there is no significant distortion of atomic densities during the formation of a diatomic molecule. Figure~\ref{fig:diatomic_densities} depicts the difference in a selection of diatomic electron densities and the corresponding sum of atomic densities, showing that, in the majority of cases, this is true.

More specifically, this assertion is true when the bonding alkali, i.e. the alkali with its electron deficient side in the bonding domain, is heavier than \ce{H}. In these cases we see the outermost electron of the other atom occupies the position of the LUMO of the bonding alkali. The orbital that this electron occupies does not appreciably change in shape, however. As a result the bond length is completely determined by the shape of the atomic species.

The exception to this are those diatomics where the bonding alkali is an \ce{H} atom. The unshielded proton of the \ce{H} atom is significantly more reactive than other species, an effect we also see when looking at bond strengths.  This results in the outer electron of the other atom occupying an orbital similar to the LUMO of the \ce{H} atom, rather than the HOMO of its parent atom.  This creates a noticeable distortion of the atomic electron densities in such cases.

The fact that each electron is largely isolated within its local domain provides little opportunity for complicated interelectronic interactions. 
As a result, we find that the qualitative and quantitative effects of electron correlation are usually small. 
The correlation energy in 1D constitutes a much smaller fraction (typically less than 0.1\%) of the total energy than in 3D.  Moreover, it largely cancels between reactants and products so that correlated bond energies are typically within 1 m\Eh\ of their uncorrelated values.  Correlated bond lengths are also similar to uncorrelated ones, especially in relative terms. 
Accordingly, we use HF structures henceforth.

\subsection{\label{subsec:triatomics} Triatomics}
We also undertook a systematic search for stable triatomic molecules, examining all possible electronic configurations generated by ground-state atoms up to, and including, carbon.  Many stable species emerge, and we report bond lengths, total energies, atomisation energies and bond energies for some of these in Table~\ref{tab:triatomics}. For the reasons discussed above, we report atomisation and bond energies only at the HF level.

In our earlier exploration\cite{ball15} of 1D reactivity, we concluded from a small set of atomisation energies that the bonds in a triatomic ABC are similar in strength to those in AB and BC.  We argued that the small deviations could be rationalised by considering the A--C dipole interaction.

The results in Table \ref{tab:triatomics} largely support this view.
\alert{For example, the H--Li and Li--Li bond strengths in \ce{H2Li3Li2} are 39 and 52 m\Eh, which are slightly higher than those in \ce{H2Li2} (36 m\Eh) and \ce{_1Li3Li2} (49 m\Eh), and this increased stability can be ascribed to the favourable dipole alignment. }
In contrast, the Li--H and H--B bond strengths in \ce{_2Li2H2B3} fall from 36 and 100 m\Eh\ to 20 and 84 m\Eh, respectively, because the boron dipole is opposed to those of the lithium and hydrogen atoms. 
The \ce{_3B5C5B3} molecule also has opposed dipoles and the B--C bond energy drops from 9 m\Eh\ in the diatomic to 7 m\Eh\ in the triatomic.

We have also found two classes of triatomic that one might have expected to be unstable. 
The first consists of a noble flanked by two alkalis with aligned dipoles (e.g.~\ce{H2He3B3} and \ce{_2Li2He4B2}) and the second consists of two nobles on one side of an alkali (e.g.~\ce{_1He2He2Li2} and \ce{_3B3He4C3}). 
Such molecules contain bonded atoms that do not form stable diatomics, e.g.~the \ce{H2He1} moiety in \ce{H2He3B3} or the \ce{_1He4C3} moiety in \ce{_3B3He4C3}.

\begin{figure}
	\includegraphics[width=0.45\textwidth]{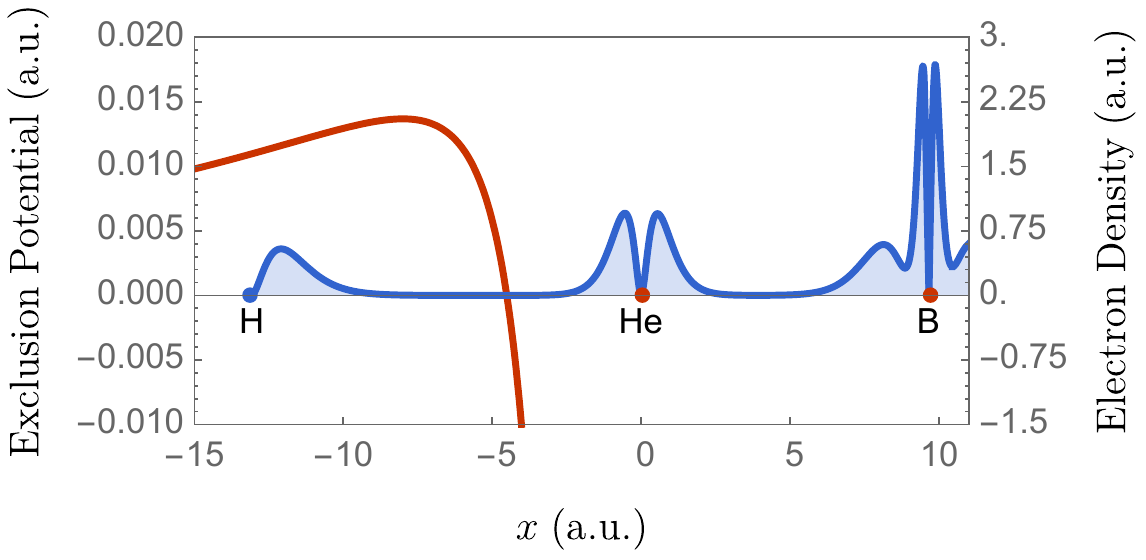}
	\includegraphics[width=0.45\textwidth]{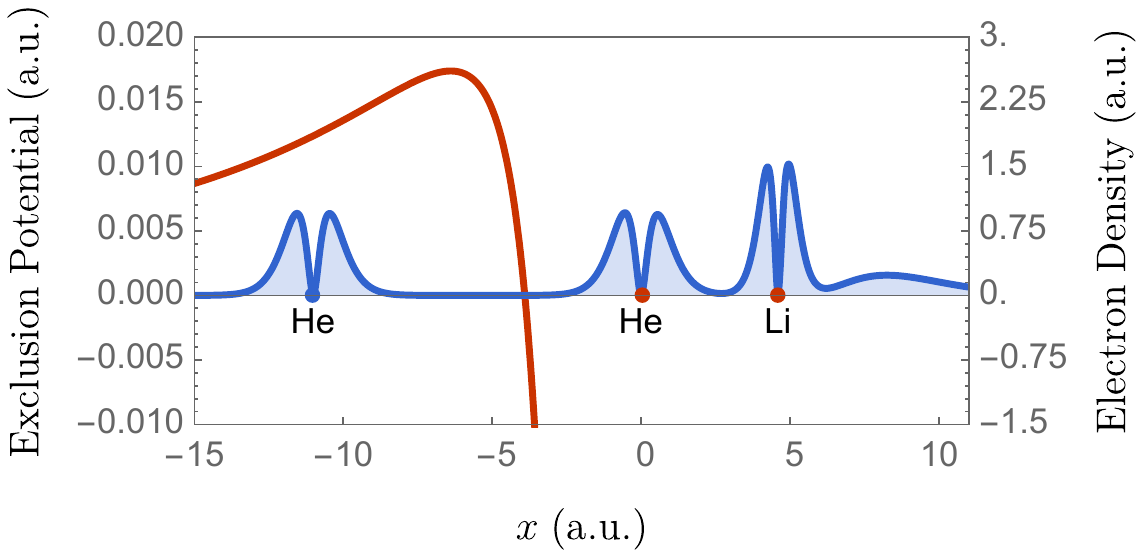}
	\caption{
	\label{fig:1he2he2li2_h2he3b3_potential}
	Left exclusion potential (red) of \ce{_1He3B3} (left) and \ce{_1He2Li2} (right) and electron density (blue) of \ce{H2He3B3} (left) and \ce{_1He2He2Li2} (right).}
\end{figure}

The exclusion potentials in Fig.~\ref{fig:1he2he2li2_h2he3b3_potential} show the attractive force which binds these unusual ABC triatomics. 
In each case, the diatomic fragment BC generates a small positive potential in its left domain which can then interact favourably with the valence electron of A.

\begingroup
\squeezetable
\begin{table*}
\caption{
\label{tab:triatomics}
Equilibrium bond lengths (bohr), total energies (\Eh), HF atomisation energies (\Eatom, m\Eh) and HF bond dissociation energies (\EAB\ and \EBC, m\Eh) of triatomic molecules.}
\begin{ruledtabular}
\begin{tabular}{@{}l *{2}{D{.}{.}{2.3}} *{3}{D{.}{.}{2.6}} *{1}{D{.}{.}{2.3}} *{2}{D{.}{.}{1.3}}}
	\cc{Molecule}              & \mc{2}{c}{Bond length} & \mc{3}{c}{Total energy} &	&	&	\\
	                 \cline{2-3} \cline{4-6}
	\cc{ABC}			&	\cc{\RAB}	&	\cc{\RBC}	&	\cc{--\EHF}	&	\cc{--\EMP}	& \cc{--\EMPP}	&	\cc{\Eatom}	&	\cc{\EAB}	&	\cc{\EBC}	\\
	\colrule
	\ \,\ce{H2H2B3}		&	17.620		&	3.296		&	26.458006	&	26.468322	&	26.469874	&	100.497		&	0.404		&	\dfoot		\\ 
	\ce{_1H1H3B2}		&	2.795		&	8.942		&	26.735055	&	26.74627	&	26.74818	&	\dcolcolor{red}310.500		&	215.008		&	\dcolcolor{red}89.366	\\ 
	\ \,\ce{H2He3B3}	&	13.090		&	9.665		&	29.101690	&	29.11340	&	29.11529	&	1.260		&	0.799		&	\dfoot		\\ 
	\ce{_1H1He4C3}		&	2.025		&	16.294		&	41.972376	&	41.9891		&	41.9917		&	139.071		&	\dfoot		&	1.680		\\ 
	\ \,\ce{H2Li3Li2}	&	5.336		&	8.860		&	16.604027	&	16.611622	&	16.61313	&	88.514		&	39.380		&	52.107		\\
	\ce{_1H2Li3Be2}		&	5.243		&	7.048		&	24.126866	&	24.13849	&	24.14078	&	203.198		&	174.387		&	29.172		\\
	\ce{_1H2Li4B2}		&	5.387		&	13.356		&	34.094432	&	34.1091		&	34.11168	&	229.167		&	175.556		&	55.142		\\
	\ce{_1H2Be3Li2}		&	3.946		&	7.047		&	24.111979	&	24.12320	&	24.12529	&	188.310		&	159.500		&	24.674		\\
	\ \,\ce{H3B4Li2}	&	10.289		&	13.443		&	33.925825	&	33.9393		&	33.9414		&	60.560		&	6.949		&	54.178		\\
	\ce{_1H3B4Li1}		&	9.099		&	14.048		&	34.040693	&	34.0553		&	34.0579		&	\dcolcolor{red}108.382		&	\dcolcolor{red}82.704		&	12.890		\\
	\ \,\ce{H3B5B3}		&	10.281		&	19.510		&	51.24102	&	51.2605		&	51.2631		&	26.006		&	7.115		&	19.62		\\
	\ce{_1H3B5B2}		&	9.205		&	19.492		&	51.39508	&	51.4159		&	51.4192		&	\dcolcolor{red}113.018		&	\dcolcolor{red}74.956		&	17.53		\\
	\ce{_1H3C3H1}		&	6.649		&	6.649		&	39.422734	&	39.4393		&	39.44217	&	332.352		&	166.062		&	166.062	\\
	\ce{_1H3C5B3}		&	6.632		&	16.058		&	64.12276	&	64.1480		&	64.1517		&	174.867		&	165.657		&	8.577		\\
	\ce{_1He1H3B3}		&	2.027		&	10.290		&	29.244835	&	29.25684	&	29.25879	&	144.404		&	138.023		&	7.014		\\
	\ce{_1He2He2Li2}	&	11.009		&	4.601		&	14.503682	&	14.511311	&	14.513255	&	10.081		&	0.104		&	\dfoot		\\ 
	\ce{_1He2Li3Li2}	&	4.601		&	8.737		&	19.317942	&	19.327327	&	19.329411	&	59.507		&	10.373		&	49.530		\\
	\ce{_1He3B4Li2}		&	10.949		&	13.320		&	36.662290	&	36.6778		&	36.6805		&	54.103		&	0.492		&	53.642		\\
	\ce{_1Li2H2Li2}		&	5.210		&	5.524		&	16.749370	&	16.75602	&	16.759498	&	233.857		&	197.450		&	59.832		\\
	\ce{_1Li2H3B3}		&	5.191		&	10.017		&	34.055957	&	34.07025	&	34.07275	&	190.692		&	184.310		&	16.667		\\ 
	\ce{_2Li2H2B3}		&	5.619		&	3.338		&	33.985753	&	33.999903	&	34.002215	&	120.488		&	20.395		&	84.081		\\ 
	\ce{_2Li2H3C3}		&	5.533		&	6.789		&	46.820115	&	46.83917	&	46.84223	&	221.976		&	55.686		&	185.569	\\
	\ce{_2Li2He3Be2}	&	4.574		&	12.299		&	26.677407	&	26.68967	&	26.69200	&	10.817		&	\dfoot		&	0.840		\\ 
	\ce{_2Li2He4B2}		&	4.514		&	20.723		&	36.623610	&	36.6388		&	36.6414		&	15.423		&	\dfoot		&	5.45		\\ 
	\ce{_1Li3Li3Li2}	&	8.69		&	8.966		&	24.127495	&	24.13851	&	24.140705	&	104.226		&	55.092		&	55.092		\\
	\ce{_2Li3Li3Be2}	&	8.833		&	7.04		&	31.51126	&	31.52547	&	31.52797	&	79.832		&	51.021		&	30.698		\\ 
	\ce{_1Li3Li4B3}		&	8.715		&	14.190		&	41.440738	&	41.457727	&	41.46047	&	67.717		&	53.951		&	18.583		\\
	\ce{_2Li3Li4C3}		&	8.948		&	10.461		&	54.201813	&	54.2238		&	54.2272		&	95.918		&	53.432		&	46.784		\\
	\ce{_2Li3Be3Li2}	&	7.077		&	7.075		&	31.482743	&	31.49682	&	31.49927	&	51.318		&	22.507		&	22.507		\\
	\ce{_1Li4B5B3}		&	13.937		&	19.704		&	58.75720	&	58.7802		&	58.7835		&	34.426		&	15.54		&	20.66		\\
	\ce{_2Be2H3B3}		&	3.983		&	10.32		&	41.445882	&	41.46312	&	41.46577	&	172.461		&	\dcolcolor{red}166.08		&	8.825		\\
	\ce{_2Be4B5B3}		&	12.580		&	19.444		&	66.15358	&	66.1798		&	66.1834		&	22.652		&	3.761		&	19.19		\\
	\ce{_2B4H2B3}		&	25.12		&	3.288		&	51.3170		&	51.3367		&	51.3397		&	10.198		&	1.9			&	\dfoot		\\ 
	\ce{_3B3H3C3}		&	10.164		&	6.747		&	64.126951	&	64.15210	&	64.15578	&	179.060		&	12.770		&	172.678	\\
	\ce{_3B3He4C3}		&	9.794		&	18.02		&	66.69221	&	66.7183		&	66.7221		&	1.39		&	\dfoot		&	0.93		\\ 
	\ce{_2B5Be4B3}		&	30.000		&	12.358		&	66.1349		&	66.1608		&	66.1644		&	4.00		&	0.5			&	\dfoot		\\ 
	\ce{_3B5C5B3}		&	16.126		&	16.129		&	88.82132	&	88.8553		&	88.8597		&	\dcolcolor{red}15.916		&	6.71		&	6.71		\\ 
\end{tabular}
\end{ruledtabular}
\footnotetext[1]{After breaking this bond, the remaining diatomic is unstable.}
\end{table*}
\endgroup

\subsection{\label{subsec:tetraatomics} Tetra-atomics}
Table~\ref{tab:tetraatomics} presents results for the stable tetra-atomic molecules formed by dimerising \ce{_1H3B2}, \ce{_1H2B3} and \ce{H3B3}.  These results confirm that the length and strength of a bond are largely independent of its environment but, as in the triatomic study, we find significant increases in some bond strengths as a result of favorable dipole interactions. 
For example, in \ce{_2B3H1H3B3}, the central \ce{H-H} bond is approximately 50 m\Eh\ stronger than that in \ce{H1H1}, and the right bond is almost four times as strong as in \ce{H3B3}.
However, this effect is not universal.  
\alert{For example, the dipoles in \ce{_1H3B5B3H} are all aligned, yet the individual bonds are not strengthened and, indeed, the central bond is weaker than in the corresponding diatomic. 
This is because the central domain houses five electrons, forcing the bond to be long ($>$ 18 bohrs) and greatly reducing the dipole stabilisation.}

In most of the stable species ABCD, the central pair BC is also a stable diatomic.  However, this is not the case in \ce{_3B2H4B3H} and \ce{_2B3H2H2B3}. 
In both of these, the central bond is significantly weaker than in the other tetra-atomics and they are therefore analogous to the loosely associated triatomics in Table~\ref{tab:triatomics}.


\alert{
We also found two molecules, \ce{_3B2H3B3H1} and \ce{_1H3B2H3B3}, where each individual bond is present in a stable diatomic but the overall tetra-atomic is not bound.
The exclusion potential of the fragment \ce{_3B2H1}, which is present in both \ce{_3B2H4B3H} and \ce{_3B2H3B3H1}, becomes positive beyond 8 bohr to the right of the H atom and this provides the driving force for bonding in the molecule \ce{_3B2H4B3H}.
However, \ce{_3B2H3B3H1} remains unbound because of the unfavourable dipole interactions between the left boron atom and the other atoms.
Similar dipole interactions also appear in \ce{_1H3B2H3B3}.
It is possible that the \ce{_3B2H1} and \ce{_2B3H1} fragments in \ce{_3B2H3B3H1}, or the \ce{_1H3B2H1} and \ce{_2B3} fragments in \ce{_1H3B2H3B3} associate weakly at a very large separation but that the (30,50) basis cannot adequately describe this.
}


\begingroup
\squeezetable
\begin{table*}
\caption{
\label{tab:tetraatomics}
HF equilibrium bond lengths (bohr), total energies (\Eh), HF atomisation energies (\Eatom, m\Eh) and HF bond dissociation energies (\EAB, \EBC\ and \ECD, m\Eh) of tetra-atomic molecules.
}
\begin{ruledtabular}
\begin{tabular}{@{}l *{3}{D{.}{.}{2.3}} *{3}{D{.}{.}{2.6}} *{1}{D{.}{.}{1.3}} *{3}{D{.}{.}{1.3}}}
	\cc{Molecule}		&		\mc{3}{c}{Bond length}		&			\mc{3}{c}{Total energy}				&				&				&				&				\\
										\cline{2-4}								\cline{5-7}
	\cc{ABCD}			&	\cc{AB}	&	\cc{BC}	&	\cc{CD}	&	\cc{--\EHF}	&	\cc{--\EMP}	& \cc{--\EMPP}	&  \cc{\Eatom}	&	\cc{\EAB}	&	\cc{\EBC}	&	\cc{\ECD}	\\
	\colrule
	\ce{_3B2H4B3H}		&	3.284	&	23.316	&	10.349	&	51.822463	&	51.842		&	51.845		&	107.48		&	\dfoot		&	1.0			&	5.5			\\
	\ce{_2B3H2H2B3}		&	8.880	&	18.08	&	3.290	&	51.980101	&	52.0013		&	52.0048		&	\dcolcolor{red}198.034		&	\dcolcolor{red}90.492		&	2.449		&	\dfoot		\\
	\ce{_2B3H1H3B3}		&	9.085	&	2.795	&	9.828	&	52.115557	&	52.1361		&	52.1392		&	\dcolcolor{red}333.182		&	\dcolcolor{red}75.51		&	231.308		&	22.682		\\
	\ce{_3B2H1H3B3}		&	3.356	&	2.715	&	10.35	&	51.99337	&	52.0142		&	52.0173		&	278.35		&	82.682		&	171.879		&	2.84		\\
	\ce{_1H3B5B2H1}		&	8.979	&	19.70	&	3.297	&	51.993816	&	52.015		&	52.019		&	\dcolcolor{red}211.6		&	\dcolcolor{red}77.3		&	16.1		&	98.6		\\
	\ce{_1H3B5B3H}		&	9.253	&	18.48	&	10.340	&	51.901493	&	51.923		&	51.926		&	\dcolcolor{red}119.429		&	\dcolcolor{red}73.090		&	17.56		&	6.41		\\
	\ce{_1H3B3H3B2}		&	8.880	&	9.99	&	8.88	&	52.06129	&	52.083		&	52.087		&	\dcolcolor{red}212.18		&	\dcolcolor{red}74.54		&	21.20		&	\dcolcolor{red}81.1		\\
	\ce{_1H2B3H3B3}		&	3.298	&	8.854	&	10.155	&	51.999949	&	52.0210		&	52.0244		&	\dcolcolor{red}217.728		&	99.060		&	\dcolcolor{red}88.595		&	\dcolcolor{red}21.238		\\
	\ \,\ce{H3B3H3B3}	&	10.348	&	9.128	&	9.825	&	51.907673	&	51.9280		&	51.9311		&	\dcolcolor{red}125.255		&	6.587		&	\dcolcolor{red}75.332		&	\dcolcolor{red}23.818		\\
\end{tabular}
\end{ruledtabular}
\footnotetext[1]{After breaking this bond, the remaining molecule is unstable.}
\end{table*}
\endgroup

\subsection{\label{subsec:polymers} Polymers}
In our early work on 1D chemistry, \cite{loos15} we examined the bond length and energy within the hydrogen nanowire --- an infinite chain of alternating protons and electrons --- using a periodic HF calculation. 
Using \LegLag, we can study the same system as the extrapolated limit of a sequence of finite chains and we can also examine other homogeneous, or heterogenous, polymers. 

For each polymer, we studied a range of short oligomers and fit their properties to the functions of the type $\sum_{k=0}^2 a_k n^{-k}$, where $n$ is the number of monomer units in the oligomer.
We then extrapolated these functions to the infinite polymer, i.e. $n \to \infty$.
For computational efficiency, we used the (30,30) basis set, rather than the (30,50) set used above.
Our results are summarised in Table \ref{tab:polymers} and we report only the digits that have converged as the basis set is increased to the (30,30) set.

Results for the hydrogen polymer agree with our periodic calculations \cite{loos15} and confirm that the \ce{H-H} bond becomes longer (stretching from around 2.6 bohrs to 2.8 bohrs) and stronger upon polymerisation.  The lengthening / strengthening trend is ubiquitous and results from a competition between growing numbers of repulsive interelectronic interactions (which are reduced if the polymers expand by a few percent) and an accumulation of favourable dipole interactions (which stabilise the polymer relative to the monomers).

However, not all of the polymers in Table~\ref{tab:polymers} follow this pattern and the \ce{(_1H1He1)_n} and \ce{(_1H2B3)_n} polymers are notable exceptions. 
In both of these, the inter-monomer bonds are exceptionally long and the resulting stabilisation is small.  Because these new bonds do not arise in stable diatomics, these ``polymers'' are better viewed as loose aggregates.

\begin{table*}
\caption{
\label{tab:polymers}
Equilibrium bond lengths (\RAB\ and \RBA, bohr), energies per monomer (\EHF, \Eh) and stabilisation energy per monomer ($E_{\rm stab}$, m\Eh) of 1D polymers at the HF level of theory.}
\begin{ruledtabular}
\begin{tabular}{@{}l *{1}{D{.}{.}{2.3}} *{1}{D{.}{.}{2.6}} @{} *{2}{D{.}{.}{2.2}} *{1}{D{.}{.}{2.6}} *{1}{D{.}{.}{1.3}}}
					&		\mc{2}{c}{Monomer}		&						\mc{4}{c}{Polymer}									\\
								\cline{2-3}											\cline{4-7}
	\cc{AB}			&	\cc{\RAB}	&	\cc{--\EHF}		&	\cc{\RAB}	&	\cc{\RBA}	&	\cc{--\EHF} &	\cc{$E_{\rm stab}$}	\\
	\colrule
	\ \,\ce{H1H1}	&	2.636		&	1.184572		&	2.798		&	2.798		&	1.420210	&	235.638				\\
	\ce{_1Li3Li2}	&	8.693		&	16.064647		&	9.0			&	9.0			&	16.13383	&	69.18				\\
	\ce{_1H2Li1}	&	5.152		&	8.681782		&	5.46		&	5.46		&	8.752991	&	71.209				\\
	\ce{_1H3B2}		&	8.880		&	26.020047		&	9.6			&	9.6			&	26.0474		&	27.4				\\
	\ce{_2Li4B2}	&	13.330		&	33.418876		&	14.0		&	14.0		&	33.447		&	28					\\
	\ce{_1H1He1}	&	2.025		&	3.880313		&	2.025		&	10			&	3.882261	&	1.948				\\
	\ce{_1H2B3}		&	3.298		&	25.957601		&	3.3			&	20			&	26.0		&	0					\\
\end{tabular}
\end{ruledtabular}
\end{table*}

\section{\label{sec:rules} Rules of 1D bonding}
Our studies reveal that chemistry in 1D is largely local. 
The combination of particle impenetrability and strong shielding causes distant particles to have very little effect on each another and, as a result, the functional groups in 1D chemistry are essentially the diatomic units within a molecule.  This reduction requires us to understand the bonding in diatomic molecules and has led us to three simple rules which describe all of the bound diatomics reported in Table~\ref{tab:diatomics}:
\begin{itemize}
	\item Two alkalis with aligned dipoles bind
	\item Two alkalis with unaligned dipoles bind if their nuclear charges differ by at least two
	\item A noble binds to an alkali's electron-deficient side
\end{itemize}
Strong bonds result from three ingredients: 
\begin{itemize}
	\item Light atoms
	\item Aligned atomic dipoles
	\item Low electron occupations in each domain
\end{itemize}
The first ingredient improves electron-nuclear attraction (because of reduced shielding);  the last also enhances Coulombic attraction and also reduces kinetic energy.

In general, a polyatomic is strongly bound if all of its constituent diatomics are separately stable.  There are interesting exceptions (such as the stable triatomic \ce{H2H2B3} and the unstable tetra-atomic \ce{_3B2H3B3H1}) but it is true for all of the tightly bound triatomics that we have identified.  
\alert{
Curiously, the rule incorrectly predicts \ce{_3B2H3B3H1} and \ce{_1H3B2H3B3} to be strongly bound when, in fact, it turns out that one of their constituent bonds is insufficiently strong to overcome the unfavourable dipole interactions.  
}Fortunately, this is the only example that we have found where the rule fails.

\section{\label{sec:conclusion} Conclusion}
Using our newly developed electronic structure program for 1D molecules, \LegLag, we have performed an extensive survey of 1D chemistry. 
By adopting improved basis functions, we have been able to identify and characterise a wide variety of stable molecules and a small set of polymers.
Many of these are novel structures and, prior to this work, would not have been expected to exist.

We have also developed an understanding of the bonding interactions in these molecules and we have identified the most significant factors that contribute to their stability.  This has allowed us to formulate a set of simple rules which predict whether a putative 1D molecule is stable.

\begin{acknowledgments}
C.J.B.~is grateful for an Australian Postgraduate Award.
P.F.L.~thanks the Australian Research Council for a Discovery Early Career Researcher Award (DE130101441) and a Discovery Project grant (DP140104071).
P.M.W.G.~thanks the Australian Research Council for funding (Grants No.~DP140104071 and DP160100246).
P.F.L.~and P.M.W.G.~also thank the NCI National Facility for grants of supercomputer time. 
\end{acknowledgments}

\appendix

\section{\label{app:ints} Integrals}
It is convenient to define the parity function
\begin{equation}
	\eps_n =	\begin{cases}
					1,	&	n \text{ is even}	\\
					0,	&	n \text{ is odd}
				\end{cases}
\end{equation}

\subsection{One-electron integrals}
If we assume $\mu \le \nu$, the kinetic integrals are
\begin{equation}
	(\bR_\mu | \hat{T} | \bR_\nu) = \frac{\alpha^2}{2} \left[ \frac{(2\mu)_3}{6\sqrt{(\mu)_2(\nu)_2}} - \delta_{\mu\nu} \right]
\end{equation}
and
\begin{equation}
	(\bM_\mu| \hat{T} | \bM_\nu) = \eps_{\mu+\nu} \sqrt{\frac{(\mu)_4(\mu+\frac{3}{2})(\nu+\frac{3}{2})}{(\nu)_4}} \frac{\mu^2 + 3\mu - 1}{6\om^2}
\end{equation}

The potential to the left of $\bR_\mu\bR_\nu$ is
\begin{equation}
	(\bR_\mu | \hat{V} | \bR_\nu) = \frac{2\alpha\ L_{\mu-1}^2(2t)}{\sqrt{(\mu)_2(\nu)_2}} \ (\nu+1)! \,U(\nu,-1,-2t)
\end{equation}
where $U$ is Tricomi's function.\cite{NISTbook}

The potentials to the left(+) or right(--) of $\bM_\mu \bM_\nu$ are
\begin{equation}
	(\bM_\mu | \hat{V}| \bM_\nu) = \pm \frac{2}{\om} \sqrt{\frac{(\mu+\frac{3}{2})(\nu+\frac{3}{2})}{(\mu)_4(\nu)_4}} \ P_{\mu+1}^2(z) \ Q_{\nu+1}^2(z)
\end{equation}
where $Q_m^2$ is a second-order associated Legendre function of the second kind. \cite{NISTbook}

\subsection{Clebsch-Gordan expansions}
Products of our basis functions have finite expansions
\begin{subequations}
\begin{gather}
	\bL_\mu(s) \bL_\nu(s)	= \sum_n a_n^{\mu\nu} \cL_n(s)				\\
	\bR_\mu(t) \bR_\nu(t)	= \sum_n a_n^{\mu\nu} \cR_n(t)				\\
	\bM_\mu(z) \bM_\nu(z)	= \sum_n b_n^{\mu\nu} \cM_n(z)
\end{gather}
\end{subequations}
where the expansion functions are
\begin{subequations}
\begin{align}
	\cL_n(s)	& = \frac{8\alpha}{(n)_2} \ s^2 L_{n-1}^2(2s) \exp(-2s)		\\
	\cR_n(t)	& = \frac{8\alpha}{(n)_2} \ t^2 L_{n-1}^2(2t) \exp(-2t)		\\
	\cM_n(z)	& = \frac{1}{2(n)_4\om}\ (1-z^2)\ P_{n+1}^2(z)
\end{align}
\end{subequations}
and $n$ ranges from $|\mu-\nu|+1$ to $\mu+\nu-1$.  For example,
\begin{subequations}
\begin{gather}
	\bR_2 \bR_3	= 2\sqrt{2}\ \cR_2 - 4\sqrt{2}\ \cR_3 + 5\sqrt{2}\ \cR_4	\\
	\bM_2 \bM_3	= 10\sqrt{21}\ \cM_2 - 0 \cM_3 + 35\sqrt{21}\ \cM_4
\end{gather}
\end{subequations}
We call these Clebsch-Gordan (CG) expansions and the coefficients are given by
\begin{subequations}
\begin{gather}
	a_n^{\mu\nu} = \int_0^\infty \frac{L_{\mu-1}^2(t)}{\sqrt{(\mu)_2}} \frac{L_{\nu-1}^2(t)}{\sqrt{(\nu)_2}} \frac{L_{n-1}^2(t)}{t^{-2}\exp(t)} \,dt	\\
	b_n^{\mu\nu} = \int_{-1}^1	\frac{P_{\mu+1}^2(z)}{\sqrt\frac{(\mu)_4}{\mu+3/2}} \frac{P_{\nu+1}^2(z)}{\sqrt\frac{(\nu)_4}{\nu+3/2}}
								\frac{P_{n+1}^2(z)}{\frac{1-z^2}{2n+3}} \,dz
\end{gather}
\end{subequations}

\subsection{\label{app:prop} Properties of the expansion functions}
The Laplace transforms of $\cR_n$ and $\cM_n$ are
\begin{subequations}
\begin{gather}
	\int_0^\infty \cR_n(t)	\exp(-u t)\,dt	= \frac{\alpha (u/2)^{n-1}}{(1+u/2)^{n+2}}
	\\
	\int_{-1}^1 \cM_n(z)		\exp(-u z)\,dz	= \frac{(-1)^{n+1} i_{n+1}(u)}{\om u^2}
\end{gather}
\end{subequations}
where $i_n$ is a modified spherical Bessel function. \cite{NISTbook}

The moments of $\cR_n$ and $\cM_n$ are
\begin{subequations}
\begin{gather}
	\int_0^\infty \cR_n(t)  \,t^k \,dt	= \frac{\alpha (-1)^{n+1} \, k! \, (k+2)!}{(n+1)! \, \G(2+k-n) \, 2^k}
	\\
	\int_{-1}^1 \cM_n(z)	 \,z^k \,dz	= \frac{\eps_{n+k+1} \,\G(\frac{k+1}{2}) \,\G(\frac{k+2}{2})}{8\om \,\G(\frac{k-n+3}{2}) \,\G(\frac{k+n+6}{2})}
\end{gather}
\end{subequations}
The $k$th moments of $\cR_n$ and $\cM_n$ vanish if $k < n-1$.  All of the higher moments of $\cR_n$ have the same sign.  All of the higher moments of $\cM_n$ are positive.

\begin{figure}
	\includegraphics[width=0.45\textwidth]{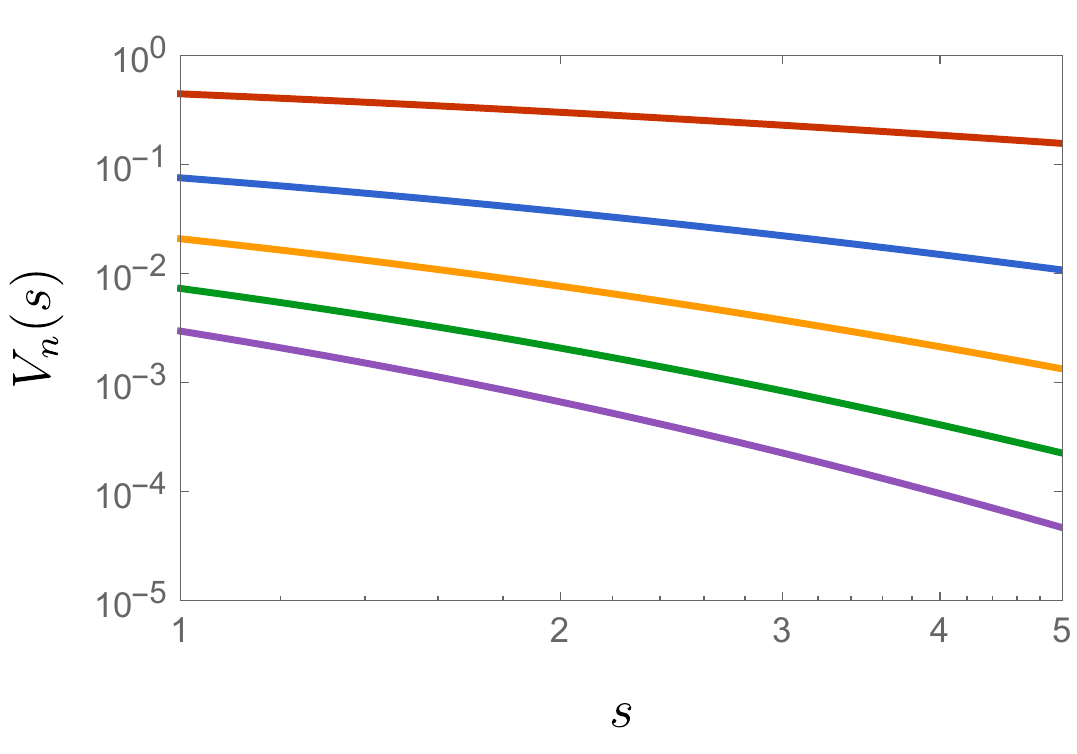}
	\includegraphics[width=0.45\textwidth]{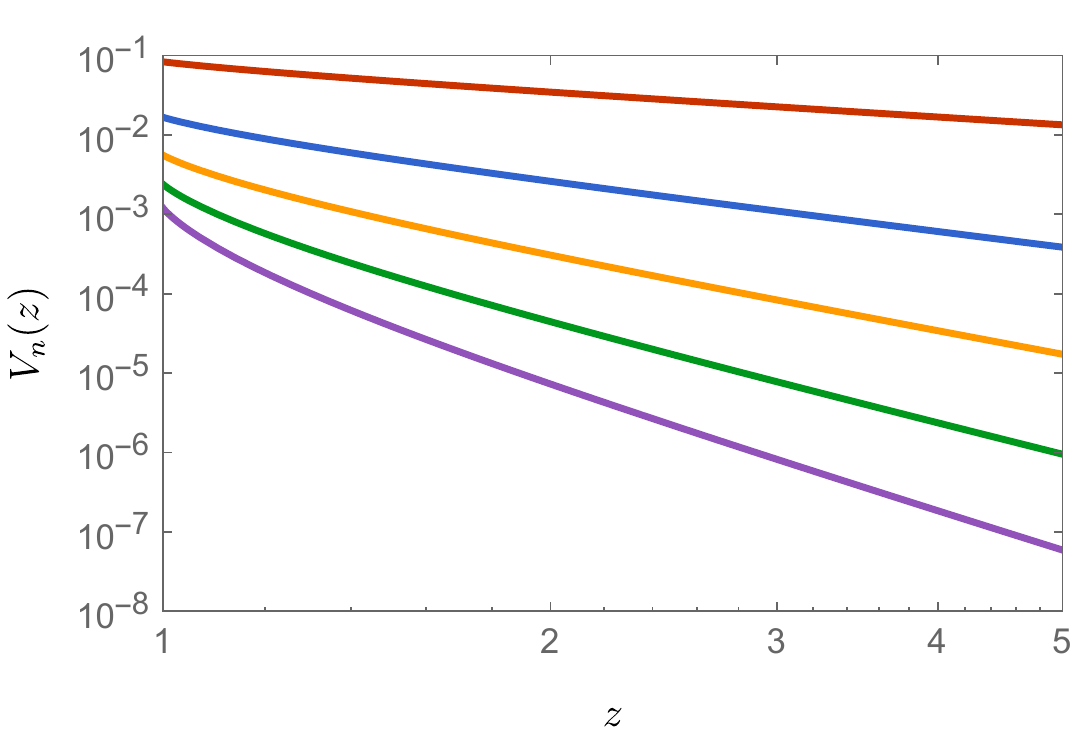}
	\caption{\label{fig:pot} Potential $V_n(x)$ to the right of $\cL_n(s)$ with $\alpha = 1$  (left) and to the right of $\cM_n(z)$ with $\om = 1$ (right).  
	From top to bottom, $n = 1,2,3,4,5$.}
\end{figure}

The potential to the left of $\cR_n$ is
\begin{equation}
	\int_0^\infty \frac{\cR_n(r)}{r-t} \,dr = 2\alpha\ \G(n)\ U(n,-1,-2t)
\end{equation}
and the potentials to the right(+) or left(--) of $\cM_n$ are
\begin{equation}
	\int_{-1}^1 \frac{\cM_n(r)}{|z-r|} \,dr = \frac{\pm\G(\frac{n}{2}) \,\G(\frac{n+1}{2})}{\G(n+\frac{5}{2}) \,8\,\om \,z^n}
										\,F\left[ \frac{n}{2},\frac{n+1}{2},n+\frac{5}{2},\frac{1}{z^2} \right]
\end{equation}
where $F$ is the Gauss hypergeometric function.\cite{NISTbook}  These potentials, which are illustrated in Fig.~\ref{fig:pot}, are monotonically decreasing and behave asymptotically as $O(x^{-n})$.

The absolute contents of $\cR_n$ and $\cM_n$ satisfy
\begin{subequations}
\begin{gather}
	\int_0^\infty		\left| \cR_n(t) \right|		\,dt < \frac{10}{9n^{5/4}}	\\
	\int_{-1}^1		\left| \cM_n(z) \right|	\,dz < \frac{1}{10n^2}
\end{gather}
\end{subequations}
These quantities can be used to compute simple upper bounds to the Coulomb integrals in the next Section.

\subsection{Coulomb integrals}
The CG expansions yield
\begin{subequations}
\begin{gather}
	(\bL_\mu \bL_\nu	| \bR_\lambda \bR_\sigma) = \sum_{mn} a_m^{\mu\nu} a_n^{\lambda\sigma} (\cL_m | \cR_n)		\\
	(\bL_\mu \bL_\nu	| \bM_\lambda \bM_\sigma) = \sum_{mn} a_m^{\mu\nu} b_n^{\lambda\sigma} (\cL_m | \cM_n)		\\
	(\bM_\mu \bM_\nu	| \bM_\lambda \bM_\sigma) = \sum_{mn} b_m^{\mu\nu} b_n^{\lambda\sigma} (\cM_m | \cM_n)
\end{gather}
\end{subequations}
The Coulomb integral\cite{ball15} between densities $f(x-X)$ and $g(y-Y)$ in different domains with $X \le Y$ is given by
\begin{equation}
	(f|g) = \int_0^\infty F(-u) G(u) \exp(-Ru) \,du
\end{equation}
where $F$ and $G$ are the Laplace transforms of $f$ and $g$ and $R = Y - X$.
In this way, we find that
\footnotesize
\begin{subequations}
\begin{gather}
	(\cL_m | \cR_n)	 =	2\alpha\ \G(m+n-1)\ U(m+n-1,-4,2\alpha R)																		\label{eq:LRint}		\\
	(\cL_m | \cM_n) =	-\frac{(-\alpha\,\om)^n\sqrt\pi}{4\om} \sum_{k=0}^\infty \frac{\G(m+n-1+2k)\ U(m+n-1+2k, n-2+2k, 2\alpha R)}
																					{\G(n+\frac{5}{2}+k)} \frac{(\alpha^2\om^2)^k}{k!}	\label{eq:LMint}		\\
	(\cM_m | \cM_n) =	\frac{\pi}{64R} \left(\frac{\om_1}{2R}\right)^{m-1} \left(-\frac{\om_2}{2R}\right)^{n-1}
						\sum_{k=0}^\infty \frac{\G(m+n-1+2k)}{\G(m+\frac{5}{2}+k)\G(n+\frac{5}{2}+k)}
						\left[ \frac{\om_2^2-\om_1^2}{4R^2} \right]^k
						P_k^{(m+\frac{3}{2},n+\frac{3}{2})}\left[ \frac{\om_2^2+\om_1^2}{\om_2^2-\om_1^2} \right]						\label{eq:MMint}
\end{gather}
\end{subequations}
\normalsize
where $U$ is Tricomi's function and $P_k^{(a,b)}$ is a Jacobi polynomial.\cite{NISTbook}

Each of the integrals \eqref{eq:LRint}, \eqref{eq:LMint} and \eqref{eq:MMint} is $O(1/R^{m+n-1})$ for large $R$.  Consequently, for domains that are far apart, many of the higher Coulomb integrals are negligible and can be safely neglected using the bound \eqref{eq:bound}.

\subsection{Quasi-integrals}
The CG expansions yield
\begin{subequations}
\begin{gather}
	\{\bR_\mu \bR_\nu	| \bR_\lambda \bR_\sigma\}	= \sum_{mn} a_m^{\mu\nu} a_n^{\lambda\sigma} \{\cR_m | \cR_n\}		\\
	\{\bM_\mu \bM_\nu	| \bM_\lambda \bM_\sigma\}	= \sum_{mn} b_m^{\mu\nu} b_n^{\lambda\sigma} \{\cM_m | \cM_n\}
\end{gather}
\end{subequations}
The quasi-integral\cite{ball15} between densities $f(x)$ and $g(y)$ in the same domain is given by
\begin{equation}
	\{f|g\} = -\frac{1}{2\pi} \int_{-\infty}^\infty F(-i k) G(i k) \log k^2 \,dk
\end{equation}
If we define the harmonic sum
\begin{equation}
	H_n = \sum_{k=1}^n \frac{1}{2k-1}
\end{equation}
assume $m \le n$ and define $\Delta = n - m$, we find that
\begin{equation}
	\{\cR_m | \cR_n\} =	\alpha \frac{(-1)^{n+1} \Delta!}{(n+1)!} \frac{\sqrt\pi}{2}
					 \sum_{k=0}^{\Delta/2} \frac{(2k+1)(2k+3)(H_{k+2}-H_{n-1-k})}{4^k\ \G(3/2-n+k)\ (\Delta-2k)!\ k!}
\end{equation}
and
\begin{equation}
	\{\cM_m | \cM_n\} =	\frac{\eps_{m+n}}{64\om\left( m+\frac{\Delta-1}{2} \right)_5}
							\times \begin{cases}
								+12	(\frac{7}{12} - H_{m-1} - H_{m+4}),			&	\Delta = 0		\\
								- \ 8	(\frac{2}{3} - H_{m} - H_{m+5}),		&	\Delta = 2		\\
								+\ 2	(\frac{25}{24} - H_{m+1} - H_{m+6}),	&	\Delta = 4		\\
								- 24\ /\left( \frac{\Delta}{2} - 2 \right)_5,	&	\Delta \ge 6
							\end{cases}
\end{equation}

%

\end{document}